\documentclass[twocolumn]{article}
\usepackage{amsmath}

\title{Constructing a genome assembly that has the maximum likelihood}
\author{Mohammadreza Ghodsi\\\texttt{mghodsi@gmail.com}}

\DeclareMathOperator*{\E}{E}

\begin{document}
\maketitle

\begin{abstract}
We formulate genome assembly problem as an optimization problem in which the objective function is the likelihood of the assembly given the reads.
\end{abstract}

\section{Introduction}
The likelihood of an assembly is proportional to the probability of observing the sequenced reads, if so many reads were generated from the assembled sequence according to the sequencing model\cite{medvedev2009maximum, validation_paper}.
Therefore, given a set of reads $R$, the genome assembly problem\footnote{
  For the sake of simplicity we use the a very basic sequencing model in much of this paper. In particular, we assume the read sampling is uniform, there are no sequencing errors, and the reads have equal lengths.
  These assumptions can be relaxed to some extent, without changing the nature of the problem.
} is to find a superstring\footnote{
  In practice an assembly may not include all of the reads.
  This can be handled by slightly adjusting the objective function\cite{validation_paper}.
} $A$, which maximizes the probability of observing $R$:
\begin{align}
  \label{probability_of_reads_given_assembly}
  \Pr\left[R \vert A\right] = \prod_{i \in R} \left(\frac{n_i}{L}\right)
\end{align}
where $n_i$ is the number of times read $i$ ``appears'' in $A$, and $L$ is the length of $A$.

\label{prefix_graph_properties}
Consider the prefix graph\footnote{
  The prefix graph\cite{vazirani2004approximation} has been used for finding an approximate shortest common superstring. It is similar to the string graph\cite{myers2005fragment} without any reduction or simplification.
}
of the read sequences in $R$. 
This graph has the following properties:
\begin{itemize}
\item It is directed.
\item Each read corresponds to a vertex.
\item The length of each edge (denoted by $l_{ij}$) is the length of the shortest prefix of read $i$, such that the remainder of read $i$ is a prefix of read $j$. 
  If two or more reads have the same sequence, they are connected by a zero length path.
  (Not a zero length cycle.)
\end{itemize}
This is a complete directed graph, and all vertices have self-loops.
Even if two reads do not overlap at all, there is an edge between them labeled by the sequence of the first read.
(These longer edges correspond to breaks in the assembly.
We refer to them as break edges.)
The edge lengths satisfy triangle inequality.
(However, note that the edge lengths are not a metric.
In particular, in general $l_{ij} \neq l_{ji}$, and always $l_{ii} \neq 0$.)
It can be shown that any assembly sequence that maximizes \eqref{probability_of_reads_given_assembly} corresponds to some walk in this graph\footnote{
  The corresponding sequence for a walk in the prefix graph is obtained by concatenation of the labels on the edges, and the sequence of the last vertex.
}.
We want to find such an optimal walk.

This work is very similar to the work of Varma, Ranade and Aluru\cite{varma2011improved}. However there are some important differences which are pointed out in section~\ref{authors_note}.

\section{A tractable optimization formulation}
\label{optimization_formulation}
We first express finding an optimal walk in terms of an integer programming problem.
We have to assume that the length of the genome being assembled (denoted by $L$) is known\footnote{
  In general, the length of the genome can not be estimated using random fragments only.
  However, assuming most of the genome is non-repetitive, the length of the genome can be estimated.
  There are many exceptions to this assumption; notably polyploid organisms.
}.
Let $x_{ij}$ denote the number of times a walk traverses the edge between vertices $i$ and $j$. 
We obtain the following integer optimization problem:

\begin{align}
  \max & \prod_{i \in R} \left(\frac{n_i}{L}\right) \label{product_objective}\\
  \forall i, n_i &= \sum_{j} x_{ij} \\
  L &= \sum_{i} \sum_{j} x_{ij} l_{ij} \label{constraint_length} \\
  \forall k, \sum_{j} x_{kj} &= \sum_{i} x_{ik} \label{constraint_euler} \\
  x_{ij} &\in \{0, 1, 2, \ldots\} \label{constraint_edge}
\end{align}

It is essential to understand that this integer program does not perfectly model the assembly problem. 
In particular it does not enforce global contiguity.
Therefore the optimal solution is an upper bound on the probability for the optimal assembly.
Also note that the graph induced by a solution is Eulerian (because of \eqref{constraint_euler}).
But such a graph may have more than one connected components.
Therefore, any integer solution will correspond to one or more cycles.

Solving an integer programming problem is not feasible for large instances.
Fortunately, we can transform this problem to a convex optimization problem.
Let us assume all variables are non-negative real numbers.
In this case, \eqref{constraint_edge} should become $x_{ij} \geq 0$.
Note that, in constraint \eqref{constraint_length}, since an optimal solution for any value of $L$ can be scaled to an optimal solution for all values of $L$, we can assume $L = 1$ without loss of generality.
Let $y_i = \frac{n_i}{L}$ denote the new variables corresponding to the graph vertices.
This means that $y_i = \Pr\left[i \vert A\right]$.
Furthermore we introduce new variables $z_i = \log y_i$, and modify the objective function accordingly.

The new optimization problem is:
\begin{align}
  \max &\sum_{i} z_i \label{convex_program_first}\\
  \forall i, z_i &\leq \log y_i \label{new_constraint_log} \\
  \forall i, y_i &= \sum_{j} x_{ij} \label{y_i_equation}\\
  1 &= \sum_{i} \sum_{j} x_{ij} l_{ij} \\
  \forall k \sum_{j} x_{kj} &= \sum_{i} x_{ik} \\
  x_{ij} & \geq 0 \label{convex_program_last}
\end{align}

This is a convex optimization problem.
The objective function is linear, and the feasible solution set is convex.
Because all constraints are linear, except \eqref{new_constraint_log}, which is convex, and the intersection of several convex sets (constraints) is also convex.

Given the optimal solution for this problem (which will, in general, be fractional), we can proceed in two ways. 
We can either use the fractional solution directly to answer standard queries on the assembly (section~\ref{using_an_optimal_fractional_solution_directly}).
Or, we can try to round this solution and generate a final sequence for the assembly (section~\ref{rounding}).
The rounding procedure must preserve the Eulerian property: for all vertices, in-degree must be equal to out-degree. 
Note that the resulting Eulerian graph may have many tours, all of which will have equal likelihood. 
Therefore any final solution (assembled sequence) is, by itself, only one of many possible solutions. 

\section{Constructing an assembly from an optimal fractional solution}
\label{rounding}
We will use the same names for the variables of the convex program~\eqref{convex_program_first}-\eqref{convex_program_last} and for the \emph{value} of these variables in an optimal solution. 
In order to round the solution from the convex program, we first round the values corresponding to the vertices (i.e. $y_i$s, representing in-degree = out-degree) and then find the best set of edges to support them. 
So at this point we know how many times each vertex is to be visited in the assembly (we get $y_i$s from the convex solution and round them up or down to $\lceil L y_i \rceil$ or $\lfloor L y_i \rfloor$ independently at random, and call the rounded value $n_i$, such that $\E[n_i] = L y_i$), but we don't know which edges to take (the convex solution only gives us fractional values for edges, we want integral values).
We want to select edges such that the total length of the assembly is minimized, while visiting each vertex exactly $n_i$ times (i.e. in-degree = out-degree = $n_i$).
This can be formulated as a minimum weight bipartite matching problem.
The edge weights are the edge lengths as defined in prefix graph, and we have a degree for each vertex.

We replicate vertex $i$, $n_i$ times.
Then translate the directed graph to a bipartite graph as follows:
Duplicate each vertex, put one in each of the two parts.
Then for each directed edge put an edge from its source vertex in the first part to its destination vertex in the second part of the bipartite graph.

This matching problem has an optimal (integral) solution that can be found in polynomial time.
This translates to a (possibly disconnected) set of Eulerian graphs embedded in the prefix graph.
Any Eulerian cycle (or set of cycles) has the same likelihood.

We have no bound on the approximation factor of this algorithm. 

\section{Simplifying the graph}
\label{simplifying_the_graph}
The size of the prefix graph as described above is quadratic in the number of reads\footnote{
  As a special case, a different kind of graph of linear size can be built assuming the reads have no errors whatsoever.
  This is not useful in practice, but for some theoretical amusement see section~\ref{graph_without_errors}.
}.
An optimization problem of this size is not practical.
Fortunately, we can reduce the size of the graph while ensuring that the optimal solution for the new graph is also an optimal solution for the original graph.
We will first give some theoretical justifications for a set of operations that transform any optimal solution on the prefix graph such that it avoids certain edges.
Next we will give a more practical recipe for efficiently constructing a simplified graph directly.

\subsection{Optimality-preserving transformations}
\label{transformations}
Consider the value of an optimal solution.
In the following, we show that several kinds of transformations of the graph do not change the value of the optimal solution.
Furthermore, since our transformations involve removing edges from the graph, any optimal solution in the ``simplified'' graph can be easily turned into an optimal solution for the original graph.
\begin{itemize}
  \item 
    Remove ``transitive'' edges. 
    If $l_{ij} + l_{jk} \leq l_{ik}$ then any optimal solution which uses the edge $ik$ can be transformed into a solution of greater than or equal likelihood which uses the two edges $ij$ and $jk$ instead.
    (Because, this transformation does not increase the total length  of the assembly \emph{and} does not decrease the product of $n_i$s.)

  \item 
    Unfortunately transitive edge removal does not affect the long ``break'' edges (See Section~\ref{prefix_graph_properties}) that exist between every pair of vertices that do \emph{not} have a significant overlap.
    These edges keep the graph nearly complete.
    Fortunately we can argue that most of these edges can removed if a better local path exists.
    In particular for every vertex $v$ which is part of a ``compact'' \emph{and} ``uniform'' path\footnote{
      We have a more rigorous theoretical definition and proof which is omitted here.
    }, we can remove all incoming and outgoing break edges.
    
    Once these edges are removed, we can collapse the paths that have no branches. 
    (Or not!)
    The important point is to note that there is a subtle difference between this and the traditional path collapsing heuristic\cite{myers2005fragment}.
    We do not collapse non-uniform paths, even if there are no branches due to overlapping reads in them. 
    For example, if the original genome contains two chromosomes with sequences $X$ and $XY$, consolidating $X$ and $Y$ into a single edge is not allowed.
    (A single edge forces the coverage of $X$ and $Y$ in the assembly to be the same, whereas the observed coverage of $X$ is twice that of $Y$.)
  
  \item
    The remaining ``break'' edges can be replaced by adding a dummy vertex, and rerouting every break edge through this vertex.
  In particular, we add an edge from every vertex to this special vertex.
  These edges have lengths and labels equal to that of the reads.
  We also add zero length edges from the special vertex back to all vertices.
  Note that this special vertex does not correspond to any variable in the optimization problem, but the edges incident on this vertex do.

\end{itemize}

\subsection{Efficient construction of a simplified graph}
This section looks very similar to section~\ref{transformations}.
It serves a subtly different purpose though.
Previously, we started with a complete graph (of size $O(n^2)$ for $n$ reads), and transformed it while preserving (the value of) an optimal solution (which conceptually had been found on the original graph).
In this section we try to construct an already simplified graph directly from the reads, with all intermediate steps requiring no more than close to linear memory.
The optimization problem is then solved on the resulting smaller graph.

\begin{enumerate}
\item
  Only find overlaps of certain significance.
\item
  Remove transitive edges.
\item
  Compress paths without any branches (in or out) if all the edges on the path have (approximately) the same length. 
  (But remember how many vertices were compressed into each edge.)
\item
  Simulate ``break'' edges between every pair of vertices by adding a dummy vertex as described in section~\ref{transformations}.
  This way, the number of additional edges will be linear (instead of quadratic) in terms of the number of vertices of the graph.
\end{enumerate}
This graph, on average, requires approximately $O(Ld)$ space to construct, where $L$ is the length of the genome and $d$ is depth of coverage. (i.e. the space requirement is linear in the size of the input reads.) 
Assuming the structure of the repeats in the genome is not very complex, the final graph has size proportional to $L$.

\section{Using an optimal fractional solution directly}
\label{using_an_optimal_fractional_solution_directly}
We do not need to build an assembly or even round the optimal fractional solution of the convex program~\eqref{convex_program_first}-\eqref{convex_program_last} to be able to answer some questions regarding the genome.
One of the most important queries that we can answer using the fractional solution directly is finding the probability of any given sequence being observed in the genome.

Given $X$, a fractional solution to~\eqref{convex_program_first}-\eqref{convex_program_last}, and a query sequence $s$, the probability of observing $s$ is the sum of probabilities of observing it over all walks (of length close to the length of $s$).
That is:
\begin{align*}
\Pr\left[s\vert X\right] = \sum_{w} \Pr\left[s \vert w\right] \Pr\left[w \vert X\right]
\end{align*}

This is not how we actually calculate the probability, but it is useful to separate the two contributing factors.
The probability of a walk only depends on the fractional solution.
(But not on the query sequence or the sequencing model.)

Let us first explain how to calculate the probability of a given walk.
The key intuition is to think of the fractional solution as the values of a ``flow'' in the edges of the graph.
Because of the Eulerian constraint, we are guaranteed that the incoming flow and the outgoing flow are equal for every vertex.
Then the probability of a walk is the amount of flow that exactly follows that particular path.

For example, assume that we have arrived at vertex $i$.
The probability of choosing the outgoing edge from $i$ to $j$ is:
\begin{align}
  \label{follow_probability}
  \Pr[\text{follow}(i, j)] = \frac{x_{ij}}{\sum_k x_{ik}}
\end{align}
Therefore the probability of a walk that visits $i_1$, $i_2$, \ldots, $i_n$ is:
\begin{align*}
  \Pr\left[\text{start}(i_1)\right] \prod_{k = 1}^{n - 1} \Pr\left[\text{follow}(i_k, i_{k + 1})\right]
\end{align*}

As mentioned above, we can not enumerate all walks to calculate the probability of a query sequence.
In the following, we present a dynamic programming algorithm that calculates this probability directly.
This algorithm assumes a sequencing model in which only substitution errors are allowed\footnote{
  We believe this algorithm can be extended to a sequencing model that allowed insertions and deletions as well as substitutions.
  However, this would require solving (or approximating) a big system of linear equations for each letter in the query sequence, and is considerably more complex.
}.

We define $T[x, i, j, y]$ as the probability of observing the prefix $[1 \ldots y]$ of $s$, and ending at position $x$ on the edge from $i$ to $j$.
If $x > 1$, then 
\begin{align*}
  T[x, i, j, y] & =  \Pr\left[\text{Subst}(e_{ij}[x],s[y])\right] \times \\ & T[x - 1, i, j, y - 1]
\end{align*}
For $x = 1$, the recurrence depends on the last value of all edges ending at vertex $i$:
\begin{align*}
  T[1, i, j, y] & = \Pr\left[\text{Subst}(e_{ij}[1],s[y])\right] \times \\ & \Pr\left[\text{follow}(i, j)\right] \times \\ & \sum_{k \in R} T[l_{ki}, k, i, y - 1] 
\end{align*}
where $\Pr\left[\text{follow}(i, j)\right]$ is defined by~\eqref{follow_probability}.

At the beginning, $T[x, i, j, 0]$ is initialized to $x_{ij}$ from the fractional solution.
At the end, the probability of observing the sequence $s$ is given by
\begin{align*}
  \sum_{x, i, j} T\left[x, i, j, \text{length}(s)\right]
\end{align*}

Note that we only need to care about the edges whose $x_{ij} > 0$.
Furthermore we can use techniques in section~\ref{simplifying_the_graph} to obtain a graph with fewer edges to begin with.
Finally, since the recurrence values for $y$ only depend on the values for $y - 1$ we only need to keep the last row of the dynamic programming table.
This means that in the worst case, the memory requirements would be linear in the size of the input reads.

\section{Reads with errors}
\label{reads_with_errors}
The convex optimization~\eqref{convex_program_first}-\eqref{convex_program_last} can be extended to the case of reads with errors, with some approximations and assumptions.
The construction and simplification of assembly graph is more complicated.
e.g. For each pair of reads, in the case of error-free reads, we only cared about the longest overlap between them.
In the case of reads with errors, (theoretically) we have to consider all overlaps.
Therefore there may be multiple edges (of different lengths) between a given pair of vertices.
Furthermore, since the overlaps are not perfect, each edge needs another attribute in addition to its length:
$p_{ij,o}$ the probability of observing sequence from read $j$ in the overlap $o$, by sequencing from $i$.

Note that $p_{ij,o}$ is a constant (between 0 and 1) in the optimization problem.
In fact $p_{ij,o}$ is strictly less than 1, even if the overlap is perfect.
Also, assuming constant error rate, the shorter the length of the overlap, the higher $p_{ij,o}$.

Finally, we need to adjust the equation~\eqref{y_i_equation} for $y_i$ (the probability of read $i$, in the convex optimization formulation).
Note that in the case of a sequencing model with errors, a read may not be a substring of the assembly sequence exactly, but the read may still be ``observed'' (with lower probability) due to sequencing errors.
The $y_i$ taking into account the probabilities of the overlaps is
\begin{align}
  y_{i} = \sum_{j,o} p_{ij,o} x_{ij,o}
\end{align}

A caveat of the new formulation is that we can not simply apply the simplification rules in section~\ref{simplifying_the_graph}.
However, with some approximating assumptions, they can be modified to extend to the case of reads with errors.
For example, the extended and more specific transitive reduction rule is: remove the edge from reads $i$ to $k$, if $l_{ik} \geq l_{ij} + l_{jk}$ and $p_{ik} \leq p_{ij}p_{jk}$.

The path compression rule can stay relatively the same if we allow for some approximation.
Except that we have to keep track of the $p$ values for the removed edges, in addition to their lengths and their number.

Due to the transitive reduction rule being more strict than in the error-free case, we will probably not be able to simplify the graph as much as before.
(And therefore the convex optimization problem will be much larger.)
However, one could relax the $p_{ik} \leq p_{ij} p_{jk}$ condition to $p_{ik}{\left(1-\mu-d\sigma\right)}^{l_{ik}} \leq p_{ij} p_{jk}$ where $\mu$ and $\sigma$ are the mean and standard deviation of the rate of sequencing errors.
With such heuristic assumptions we should be able to simplify the graph nearly as much as in the error free case.

\bibliographystyle{plain}
\bibliography{assembly-optimization}

\begin{thebibliography}{1}

\bibitem{validation_paper}
M.~Ghodsi, C.M. Hill, I.~Astrovskaya, H.~Lin, D.~Sommer, S.~Koren, and M.~Pop.
\newblock De novo likelihood-based measures for assembly validation.
\newblock Submitted to PLoS ONE, February 2013.

\bibitem{medvedev2009maximum}
Paul Medvedev and Michael Brudno.
\newblock Maximum likelihood genome assembly.
\newblock {\em Journal of computational Biology}, 16(8):1101--1116, 2009.

\bibitem{myers2005fragment}
E.W. Myers.
\newblock The fragment assembly string graph.
\newblock {\em Bioinformatics}, 21(suppl 2):ii79--ii85, 2005.

\bibitem{varma2011improved}
Aditya Varma, Abhiram Ranade, and Srinivas Aluru.
\newblock An improved maximum likelihood formulation for accurate genome
  assembly.
\newblock In {\em Computational Advances in Bio and Medical Sciences (ICCABS),
  2011 IEEE 1st International Conference on}, pages 165--170. IEEE, 2011.

\bibitem{vazirani2004approximation}
V.V. Vazirani.
\newblock {\em Approximation algorithms}.
\newblock springer, 2004.

\end{thebibliography}

\appendix
\section{Assembly graph for reads without errors}
\label{graph_without_errors}
Graph for exact reads can be built in linear space using a suffix tree with suffix links.
For each read with sequence $s$, add two strings to a generalized suffix tree: $s\$_1$, and $s\$_2$. 
Each unique read sequence would then correspond to a particular internal node of the suffix tree.
Consider all walks in this graph, that can use tree edges or suffix links.
The length of each tree edge is the length of the sub-string corresponding to that edge, and the length of each suffix link is zero.
We want to find the circular walk of given length in the tree that maximizes (a function of) the number of times each read node is visited.

\section{Authors Note}
\label{authors_note}
Since the publication of the first version of this manuscript, we have discovered that a very similar technique has been previously published by Varma, Ranade and Aluru\cite{varma2011improved}.
Please credit that publication for the convex optimization formulation of maximum likelihood assembly.

Below, we list the main differences between this work and the work of Varma et al. We hope that some may be of additional interest.
\begin{itemize}
\item The assembly graph: 
  For the most part, we use an assembly graph very similar to string graph (which is used by Varma, et al.), with one important difference;
  In the construction of our assembly graph, to collapse two edges, it is not sufficient that they are incident on a vertex with in-degree and out-degree of one.
  Using the optimization formulation, we explain why each of the graph reduction operations used during construction of a string graph are (or are not) justified. 
  (Section~\ref{simplifying_the_graph}.)
\item Assembly length:
  We assume assembly length is known, and is a constant in the optimization.
  This is due to the fact that if $L$ was a variable in~\eqref{product_objective}-\eqref{constraint_edge}, for any solution with a particular objective value, one can construct an infinite set of solutions with the same objective value by scaling the variables $x_{i, j}$, $n_i$, and $L$.
  (Section~\ref{optimization_formulation}.)
\item Calculating probability of a query sequence using the optimal (fractional) solution:
  Given the assembly graph, and an optimal solution (which provides fractional values for the edges), we give a dynamic programming algorithm to calculate the probability of a query sequence that can span several edges of the graph.
  Note that there is no need to round the optimal solution for this algorithm.
  (Section~\ref{using_an_optimal_fractional_solution_directly}.)
\item Rounding the optimal solution, and constructing a contiguous assembly:
  We describe a randomized rounding procedure that produces an Eulerian graph.
  This procedure will further simplify the solution, and possibly discard very low probability edges and vertices.
  All Eulerian cycles of the rounded solution will have equal likelihood.
  (Section~\ref{rounding}.)
\item Errors in the reads:
  We give a short recipe for (approximately) generalizing the convex optimization formulation to a sequencing model which allows errors in the reads.
  (Section~\ref{reads_with_errors}.)
\item Using a generalized suffix tree as an assembly graph for exact reads:
  Assuming the reads are absolutely error-free, we observe that we can use a suffix tree (with suffix links) as the assembly graph.
  The suffix tree, in the worst-case has linear size and can be constructed in linear time.
  Whereas the memory and computation required to build the string graph may be quadratic, because the number of overlaps is quadratic in the worst case.
  (Appendix~\ref{graph_without_errors}.)
\end{itemize}

\end{document}